\documentclass[aps,prl,preprint,floatfix,superscriptaddress,showpacs]{revtex4}

\usepackage{graphicx}
\usepackage{dcolumn}
\usepackage{ulem}
\usepackage{soul}

\begin{document}

\newcommand{\el}{\mbox{${\rm e^{-}}$ }}
\newcommand{\ps}{\mbox{${\rm e^{+}}$ }}

\title{Time dependence of the electron and positron components of the
cosmic radiation measured by the PAMELA experiment between July $2006$
and December $2015$.}



\author{O. Adriani}
\affiliation{University of Florence, Department of Physics,  
 I-50019 Sesto Fiorentino, Florence, Italy}
\affiliation{INFN, Sezione di Florence,  
 I-50019 Sesto Fiorentino, Florence, Italy}
\author{G. C. Barbarino}
\affiliation{University of Naples ``Federico II'', Department of
Physics, I-80126 Naples, Italy}
\affiliation{INFN, Sezione di Naples,  I-80126 Naples, Italy}
\author{G. A. Bazilevskaya}
\affiliation{Lebedev Physical Institute, RU-119991
Moscow, Russia}
\author{R. Bellotti}
\affiliation{University of Bari, Department of Physics, I-70126 Bari, Italy}
\affiliation{INFN, Sezione di Bari, I-70126 Bari, Italy}
\author{M. Boezio}
\affiliation{INFN, Sezione di Trieste, I-34149
Trieste, Italy}
\author{E. A. Bogomolov}
\affiliation{Ioffe Physical Technical Institute,  RU-194021 St. 
Petersburg, Russia}
\author{M. Bongi}
\affiliation{University of Florence, Department of Physics,  
 I-50019 Sesto Fiorentino, Florence, Italy}
\affiliation{INFN, Sezione di Florence,  
 I-50019 Sesto Fiorentino, Florence, Italy}
\author{V. Bonvicini}
\affiliation{INFN, Sezione di Trieste,  I-34149
Trieste, Italy}
\author{S. Bottai}
\affiliation{INFN, Sezione di Florence,  
 I-50019 Sesto Fiorentino, Florence, Italy}
\author{A. Bruno}
\affiliation{University of Bari, Department of Physics, I-70126 Bari,
Italy} 
\affiliation{INFN, Sezione di Bari, I-70126 Bari, Italy}
\author{F. Cafagna}
\affiliation{INFN, Sezione di Bari, I-70126 Bari, Italy}
\author{D. Campana}
\affiliation{INFN, Sezione di Naples,  I-80126 Naples, Italy}
\author{P. Carlson}
\affiliation{KTH Royal Institute of Technology, Department of Physics, and the Oskar Klein Centre for
Cosmoparticle Physics, AlbaNova University Centre, SE-10691 Stockholm,
Sweden}
\author{M. Casolino}
\affiliation{INFN, Sezione di Rome ``Tor Vergata'', I-00133 Rome, Italy}
\author{G. Castellini}
\affiliation{ IFAC,  I-50019 Sesto Fiorentino,
Florence, Italy}
\author{C. De Santis}
\affiliation{INFN, Sezione di Rome ``Tor Vergata'', I-00133 Rome, Italy}
\affiliation{University of Rome ``Tor Vergata'', Department of
Physics,  I-00133 Rome, Italy}
\author{V. Di Felice}
\affiliation{INFN, Sezione di Rome ``Tor Vergata'', I-00133 Rome, Italy}
\affiliation{
Agenzia Spaziale Italiana (ASI) Science Data Center, Via
del Politecnico snc, I-00133 Rome, Italy}
\author{A. M. Galper}
\affiliation{National Research Nuclear University MEPhI,  RU-115409
Moscow, Russia}  
\author{A. V. Karelin}
\affiliation{National Research Nuclear University MEPhI,  RU-115409
Moscow, Russia}
\author{S. V. Koldashov}
\affiliation{National Research Nuclear University MEPhI,  RU-115409
Moscow, Russia}  
\author{S. A. Koldobskiy}
\affiliation{National Research Nuclear University MEPhI,  RU-115409
Moscow, Russia}  
\author{S. Y. Krutkov}
\affiliation{Ioffe Physical Technical Institute,  RU-194021 St. 
Petersburg, Russia}
\author{A. N. Kvashnin}
\affiliation{Lebedev Physical Institute, RU-119991
Moscow, Russia}
\author{A. Leonov}
\affiliation{National Research Nuclear University MEPhI,  RU-115409
Moscow, Russia}  
\author{V. Malakhov}
\affiliation{National Research Nuclear University MEPhI,  RU-115409
Moscow, Russia}  
\author{L. Marcelli}
\affiliation{University of Rome ``Tor Vergata'', Department of
Physics,  I-00133 Rome, Italy}
\author{M. Martucci}
\affiliation{University of Rome ``Tor Vergata'', Department of
Physics,  I-00133 Rome, Italy} 
\affiliation{INFN, Laboratori Nazionali di Frascati, Via Enrico Fermi 40,
I-00044 Frascati, Italy}
\author{A. G. Mayorov}
\affiliation{National Research Nuclear University MEPhI,  RU-115409
Moscow, Russia}
\author{W. Menn}
\affiliation{Universit\"{a}t Siegen, Department of Physics,
D-57068 Siegen, Germany}
\author{M. Merg\'{e}}
\affiliation{INFN, Sezione di Rome ``Tor Vergata'', I-00133 Rome, Italy}
\affiliation{University of Rome ``Tor Vergata'', Department of
Physics,  I-00133 Rome, Italy} 
\author{V. V. Mikhailov}
\affiliation{National Research Nuclear University MEPhI,  RU-115409
Moscow, Russia}  
\author{E. Mocchiutti}
\affiliation{INFN, Sezione di Trieste,  I-34149
Trieste, Italy}
\author{A. Monaco}
\affiliation{University of Bari, Department of Physics, I-70126 Bari, Italy}
\affiliation{INFN, Sezione di Bari, I-70126 Bari, Italy}
\author{N. Mori}
\affiliation{INFN, Sezione di Florence,  
 I-50019 Sesto Fiorentino, Florence, Italy}
\author{R. Munini}
\email[Corresponding author: Riccardo Munini, email: ]{riccardo.munini@ts.infn.it}
\affiliation{INFN, Sezione di Trieste, I-34149
Trieste, Italy}
\affiliation{University of Trieste, Department of Physics, 
I-34147 Trieste, Italy}
\author{G. Osteria}
\affiliation{INFN, Sezione di Naples,  I-80126 Naples, Italy}
\author{B. Panico}
\affiliation{INFN, Sezione di Naples,  I-80126 Naples, Italy}
\author{P. Papini}
\affiliation{INFN, Sezione di Florence,  
 I-50019 Sesto Fiorentino, Florence, Italy}
\author{M. Pearce}
\affiliation{KTH Royal Institute of Technology, Department of Physics, and the Oskar Klein Centre for
Cosmoparticle Physics, AlbaNova University Centre, SE-10691 Stockholm,
Sweden}
\author{P. Picozza}
\affiliation{INFN, Sezione di Rome ``Tor Vergata'', I-00133 Rome, Italy}
\affiliation{University of Rome ``Tor Vergata'', Department of
Physics,  I-00133 Rome, Italy} 
\author{M. Ricci}
\affiliation{INFN, Laboratori Nazionali di Frascati, Via Enrico Fermi 40,
I-00044 Frascati, Italy}
\author{S. B. Ricciarini}
\affiliation{INFN, Sezione di Florence, 
 I-50019 Sesto Fiorentino, Florence, Italy}
\author{M. Simon}
\affiliation{Universit\"{a}t Siegen, Department of Physics,
D-57068 Siegen, Germany}
\author{R. Sparvoli}
\affiliation{INFN, Sezione di Rome ``Tor Vergata'', I-00133 Rome, Italy}
\affiliation{University of Rome ``Tor Vergata'', Department of
Physics,  I-00133 Rome, Italy} 
\author{P. Spillantini}
\affiliation{National Research Nuclear University MEPhI,  RU-115409
Moscow, Russia}
\author{Y. I. Stozhkov}
\affiliation{Lebedev Physical Institute, RU-119991
Moscow, Russia}
\author{A. Vacchi}
\affiliation{INFN, Sezione di Trieste,  I-34149
Trieste, Italy}
\affiliation{University of Udine, Department of Mathematics and Informatics, I-33100 Udine, Italy}
\author{E. Vannuccini}
\affiliation{INFN, Sezione di Florence, 
 I-50019 Sesto Fiorentino, Florence, Italy}
\author{G. I. Vasilyev}
\affiliation{Ioffe Physical Technical Institute, RU-194021 St. 
Petersburg, Russia}
\author{S. A. Voronov}
\affiliation{National Research Nuclear University MEPhI,  RU-115409
Moscow, Russia}
\author{Y. T. Yurkin}
\affiliation{National Research Nuclear University MEPhI,  RU-115409
Moscow, Russia}  
\author{G. Zampa}
\affiliation{INFN, Sezione di Trieste,  I-34149
Trieste, Italy}
\author{N. Zampa}
\affiliation{INFN, Sezione di Trieste,  I-34149
Trieste, Italy}
\author{M. S. Potgieter}
\affiliation{Centre for Space Research, North-West University, 2520 Potchefstroom, South Africa}
\author{E. E. Vos}
\affiliation{Centre for Space Research, North-West University, 2520 Potchefstroom, South Africa}


\date{\today}

\begin{abstract}

Cosmic-ray electrons and positrons are a unique probe of the 
propagation of cosmic rays as well as of the nature
and distribution of particle sources in our Galaxy. Recent
measurements of these particles are challenging our basic
understanding of the 
mechanisms of production,
acceleration and propagation of cosmic rays. Particularly striking are
the differences between the low energy results collected by the
space-borne PAMELA and AMS-02 experiments 
and older measurements pointing to
sign-charge dependence of the solar modulation of cosmic-ray spectra. 
The PAMELA experiment has been measuring  
the time variation of the positron and electron intensity at Earth from July 
$2006$ to December $2015$ covering the period for the  
minimum of solar cycle $23$ (2006-2009) till the middle of the maximum
of solar 
cycle $24$, through  the polarity reversal of the heliospheric magnetic 
field which took place between $2013$ and $2014$. 
The  positron to electron ratio measured in this time period clearly
shows a sign-charge dependence of the solar modulation introduced  
by particle drifts. These results provide the first clear and continuous observation  
of how drift effects on solar modulation have  unfolded
with time 
from solar minimum to solar maximum and their dependence on
the particle  
rigidity and the cyclic polarity of the solar magnetic field. 

\end{abstract}


\maketitle


\paragraph{Introduction.}
\label{Section 1}
Electrons and positrons are a natural component of the cosmic radiation. Both cosmic-ray electrons and positrons are produced in 
the interactions between cosmic-ray nuclei and the interstellar matter. Additionally, since the observed electron flux is about an 
order of magnitude larger than the positron one (e.g. \cite{des64}), a majority of electrons must be of primary origin, probably accelerated to high 
energy by astrophysical shocks generated at sites like supernova remnants (e.g. \cite{all97}).

The recent results on the positron fraction measured by PAMELA 
\cite{adr09,adr10,adr13a}, Fermi \cite{ack12}
and AMS-02 \cite{agu13,acc14} elicited an enormous interest because of the
significant discrepancy with the expected secondary behavior
(e.g. \cite{str98}) of this fraction with energy. While
most of the excitement was due to the high energy ($> 10$ GeV)
results and their connection with possible new sources, such as pulsar
(e.g. \cite{ato95,hoo09}) or dark
matter particles (e.g. \cite{tyl89,cir08,cho08}), the differences at
low energies 
also attracted considerable interest. 
These differences were particularly intriguing because previous measurements \cite{bar97,boe00,alc00b}, which were both statistical and systematical significant,
agreed at low energies ($< 5$ GeV) with the theoretical modelling (e.g. \cite{str98,del09,dib13}). This discrepancy was explained as 
an effect of charge-sign dependence of the solar modulation (e.g. \cite{pot04,mac13}), since these older measurements were taken during the 90's, 
i.e. in a period of opposite polarity of the heliospheric magnetic field (HMF) with respect to PAMELA results.



\begin{figure}[h]
\begin{center}
\includegraphics[width=13cm]{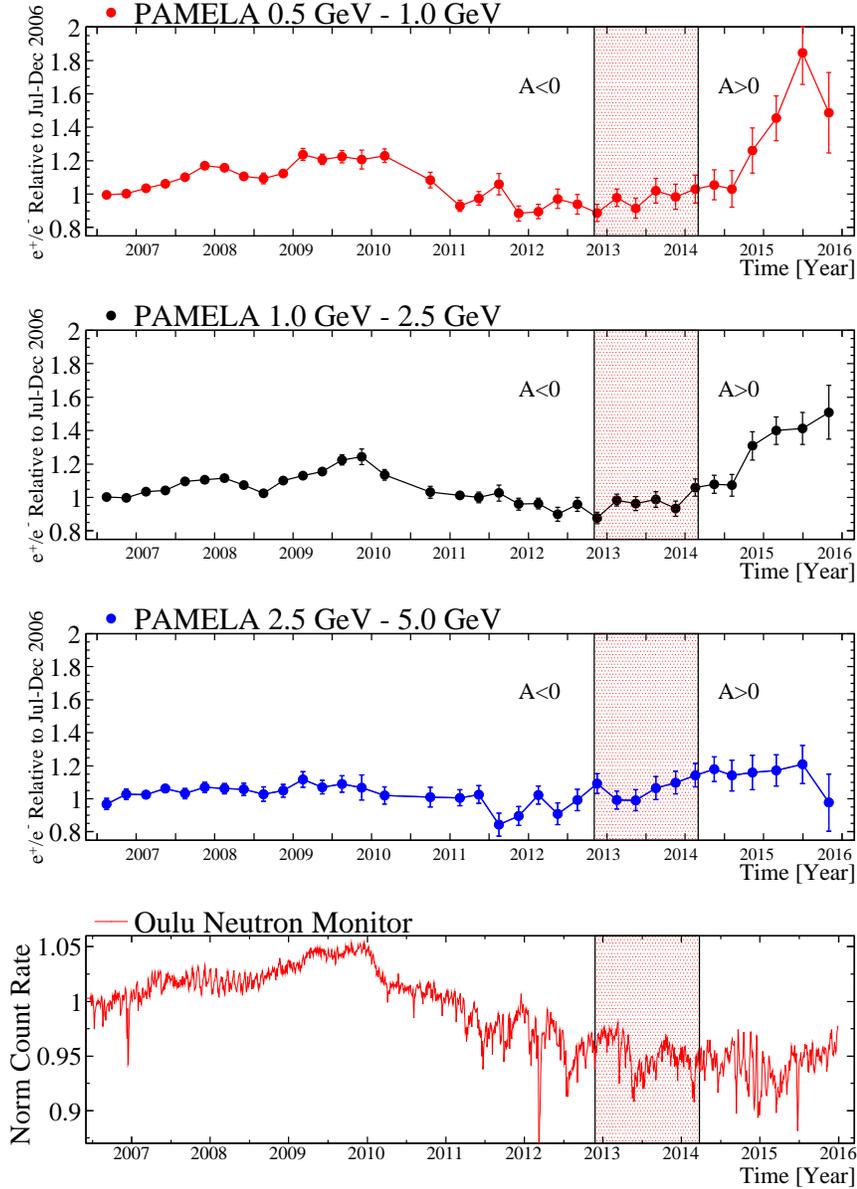}
\end{center}
\caption{The positron to electron ratios relative to July - December $2006$ measured at Earth by the
PAMELA experiment for three different energy intervals. The colored lines provide connection among the points. 
Data were selected on three month time intervals 
between July $2006$ and December $2015$.  For $2010$ only two time
intervals  were considered since the instrument  
was switched off from April to August. The shaded area corresponds to
the period with no well defined 
HMF polarity \cite{sun15}. The bottom panel shows the Oulu neutron monitor count rate
(data taken  from http://cosmicrays.oulu.fi/). 
Data are normalized to July $2006$.  
\label{f1}
} 
\end{figure}

Traversing the heliosphere, 
galactic cosmic rays (CRs) are scattered by the irregularities of the
turbulent HMF
embedded into the solar wind and undergo convection and adiabatic
deceleration  
in the expanding solar wind. As a consequence, the intensity of CRs at
Earth decreases 
with respect to the local interstellar spectrum 
\cite{pot13a}. Solar modulation has large effects 
on low energy CRs (less than a few GeV) and has negligible effects
above energies of a   
few tens of GeV. Moreover, due to the $11$-year solar activity cycle,
the intensity  
of CRs inside the heliosphere changes with time. During solar minimum
periods, the  
intensity of CRs is higher with respect to periods of solar maximum. 
This feature is well represented in the bottom panel of Figure \ref{f1},
where the counting rate of the Oulu neutron  
monitor between July $2006$ and the  end of $2015$ is shown (data are
normalized to July $2006$). 
This quantity describes well the time variations of the CR 
intensity at Earth since the neutrons are produced by the interaction
of CRs with the atmosphere and the apparatus.  

On top of the time dependence, 
a charge sign dependence of the solar modulation is expected.
The gradients and curvatures present in the HMF induce drift
motions that depend on  
the particle charge sign. During so-called 
A $<$ 0 \footnote{In the complex
sun magnetic field 
the dipole term nearly always dominates the magnetic field of the solar wind. 
A is defined as the projection of this dipole on the solar rotation
axis.}
polarity cycles such as solar cycle $23$, when the heliospheric
magnetic field is directed  
toward the Sun in the northern hemisphere, negatively 
charge particles undergo drift motion from the polar to the equatorial regions
and outwards along the 
heliospheric current sheet. Positively charged particles drift in
opposite  
directions. The situation reverses when the solar magnetic field
changes its polarity at each  
solar maximum. 
Drift effects are expected to be particularly important 
during periods of minimum solar activity and have less impact during
solar maximum \citep{fer04}. 
Indeed, solar minimum activity is the ideal condition to study the
global modulation 
processes that affect the 
CR propagation inside the heliosphere 
because very few solar-created transients disturb the modulation region.
The coincidental study of positively and negatively charged
particles allows to understand the contribution of drift motion to the
propagation of  
CRs. Furthermore, extending these measurements to solar maximum
conditions and reversal of the magnetic field polarity allows to study
how drift effects evolve with solar activity and if they actually
account for the differences in the experimental results.

In addition to the
positron fraction results discussed above,  
charge-sign effects were invoked to explain the electron  (e$^-$ +
e$^+$) and proton measurements from a few hundred MeV up to the GeV
region by 
the KET instrument on board the Ulysses spacecraft \cite{heb06} 
that explored the high latitude regions of the inner heliosphere from
$1990$ to $2009$ and the antiproton results by the BESS experiment
\cite{asa02}. Clem et al. 
\cite{cle09} reported a world summary of the positron abundance
measurements as a function of energy for different epochs of solar
magnetic polarity together. All these results point at charge sign
dependence of the  
solar modulation but are affected by large statistical and systematic
uncertainties. 
A precise understanding of the effects of solar modulation, which
significantly affects the cosmic-ray particle spectra below a few GeV, is
fundamental to fully exploit the precise experimental data
available nowadays. Low 
energy positron data, dominated by the contribution of secondary
particles, can be used to constrain propagation models and have a strong
impact on indirect dark matter searches
(e.g. \cite{lav14}). Similarly, low energy antiproton data can be used
to test 
models like annihilation or decay of dark matter particles
(e.g. \cite{gie15}) and evaporation of primordial black holes
(e.g. \cite{abe11}). Furthermore, the experimental and theoretical
investigation of the heliosphere provides information that can be
easily applied to larger astrophysical systems (e.g. \cite{sch15}).

 
PAMELA (Payload for Antimatter Matter Exploration and Light-nuclei
Astrophysics) is a satellite-borne  
experiment  \cite{pic07,boe09} designed to make long duration
measurements of the 
cosmic radiation. Results on the effects of the solar modulation on
the energy spectra of  
galactic cosmic-ray protons~\cite{adr13b} and electrons \cite{adr15}
for the 23rd solar cycle minimum (July 2006-December  
2009) have already been published by the PAMELA collaboration. 
In this paper we present a comprehensive study 
on the long-term variation of the low energy cosmic-ray positron
fraction and of 
the cosmic-ray positron to electron ratio between 500 MeV and 5 GeV
from July  
2006 to December 2015 covering the period for the  
solar minimum till the middle of the maximum of solar
cycle 24, through  the polarity reversal of the heliospheric magnetic 
field which took place between 2013 and 2014. 
The process of polar field reversal is relatively slow,
north-south asymmetric, and episodic. Sun et al. \cite{sun15}
estimated that the 
global axial dipole changed sign in October 2013; 
the northern and southern polar fields reversed in November 2012 and
March 2014, respectively, about 16 months apart.

The analysis presented in this paper is the first extensive study of CR modulation during an
unusual period of solar 
activity. It was expected that the increase in the activity 
for the $24$th solar
cycle would begin early in 2008. Instead solar minimum modulation
conditions continued until 
the end of $2009$ when the largest fluxes of galactic cosmic rays since the
beginning of the space age were recorded \cite{pot13b,str14}. The
subsequent 
maximum condition  
of solar cycle 24 continues to be unusual, with the
lowest recorded sunspot activity since accurate records began in 1750.

\paragraph{PAMELA instrument and data analysis.	}
\label{Section 2}
The PAMELA experiment was launched on June  $15^{th}$ 2006 from the
Bajkonur cosmodrome on-board the Resurs DK1 satellite and,  
since then, it has been almost continuously 
taking data. 


The apparatus comprises the following subdetectors
(from top to bottom): a
Time-of-Flight (ToF) system;  
a magnetic spectrometer; an anticoincidence system; an electromagnetic
imaging calorimeter; a shower tail catcher scintillator  and a
neutron 
detector.
A detailed description of the instruments
and data handling can be found in \cite{pic07,adr14}. 

To select a clean sample of low energy electrons and positrons, a
first selection on the goodness of the reconstructed track,  
expressed in terms 
of the  $\chi^2$ of the fit, was made. Only single track events were
selected.   
Furthermore the track was required to be reconstructed inside a
fiducial volume bounded $0.15$ cm from  
the magnet cavity walls to increase the spectrometer performance. 
The ionization losses in the ToF scintillators and in the
silicon tracker layers 
were used to  
select minimum ionizing singly charged particles. 
Albedo particles were rejected 
using the ToF velocity information. Reentrant albedo particles
were rejected comparing the 
particle rigidity with the vertical cutoff corresponding  
to the PAMELA orbital position. Only events with a measured rigidity
greater than $1.3$ times the vertical cutoff were selected.  

In the energy range between 500 MeV and 5 GeV the major source of
contamination for positrons  
is represented by protons. The positron to proton ratio is about
$10^{-3}$.  
On the other hand, antiprotons account just for a few percent of the
electron signal. Another important source of  
contamination for both electrons and positrons is represented by pions
which are created locally by the interaction  
of primary protons and nuclei 
with the PAMELA structure or pressure vessel. According to simulations
this background reaches a maximum value around $300$ - $500$ MeV  
and rapidly decreases with energy, becoming negligible above a few
GeV. Around $400$ MeV the pion background is 
about two times the positron signal and  $\sim 30-40 \%$ of the
electron signal. Finally, a non-negligible fraction of 
high rigidity ($> 10$ GV) protons are reconstructed as 
low rigidity ($<1$ GV) positively or negatively charged
particles. This is due to the presence of spurious hits in the tracker
planes 
which cause a wrong curvature reconstruction  of the track. 
These events are significant at energies below $1$ GeV and amount to a
few percent of the electron and positron signal. 
All these hadron background were rejected using a combination of
calorimeter variables defined in order to emphasize the different 
topological development of  
the electromagnetic and hadronic shower inside the PAMELA
calorimeter. For more details on the analysis see
\citep{adr13a,adr15,mun16}. 

For this study, galactic positrons and electrons were selected 
between 0.5 and 5 GeV. 
The positron to electron ratio was measured on three-month time
periods between July $2006$ and December $2015$.   
This energy and time division was chosen as the best balance between
the statistics, the energy resolution and the time resolution. 
A total of $35$ time intervals were obtain.  
For $2010$ only two time intervals  were considered since the instrument 
was switched off from April to August because of satellite problems. 

\paragraph{Results.}
Figure \ref{f1} shows the results on the time dependence of the
positron to electron ratio.  
Each of the three panels represents a different energy interval. Data 
were normalized to the values measured between July and December $2006$. As can be noticed, the statistical errors
on the positron to electron ratio increase with time. This  
 decrease in statistics was due to a reduction in the tracker
efficiency with time \cite{adr15}. The red shaded area represents  
 the time interval during which the process of polar field reversal
took place. 

The results clearly show a time dependence of the positron to electron ratio. 
In the first two energy intervals of Figure \ref{f1} ($0.5$ - $1$
and $1$ - $2.5$ GeV) 
an increase of the ratio was observed up to the end of $2009$. During
this time period  
positrons at Earth increased about $20 \%$ more than electrons. For the
third energy interval ($2.5$ - $5.0$ GeV) this    
increase was $\sim 10\%$.  
From 
Figure \ref{f1}, bottom panel, it can be noticed 
that minimum modulation was reached at the end of $2009$, when the
neutron monitor count rate reached 
its maximum values. 
\begin{figure}[h]
\begin{center}
\includegraphics[width=\columnwidth]{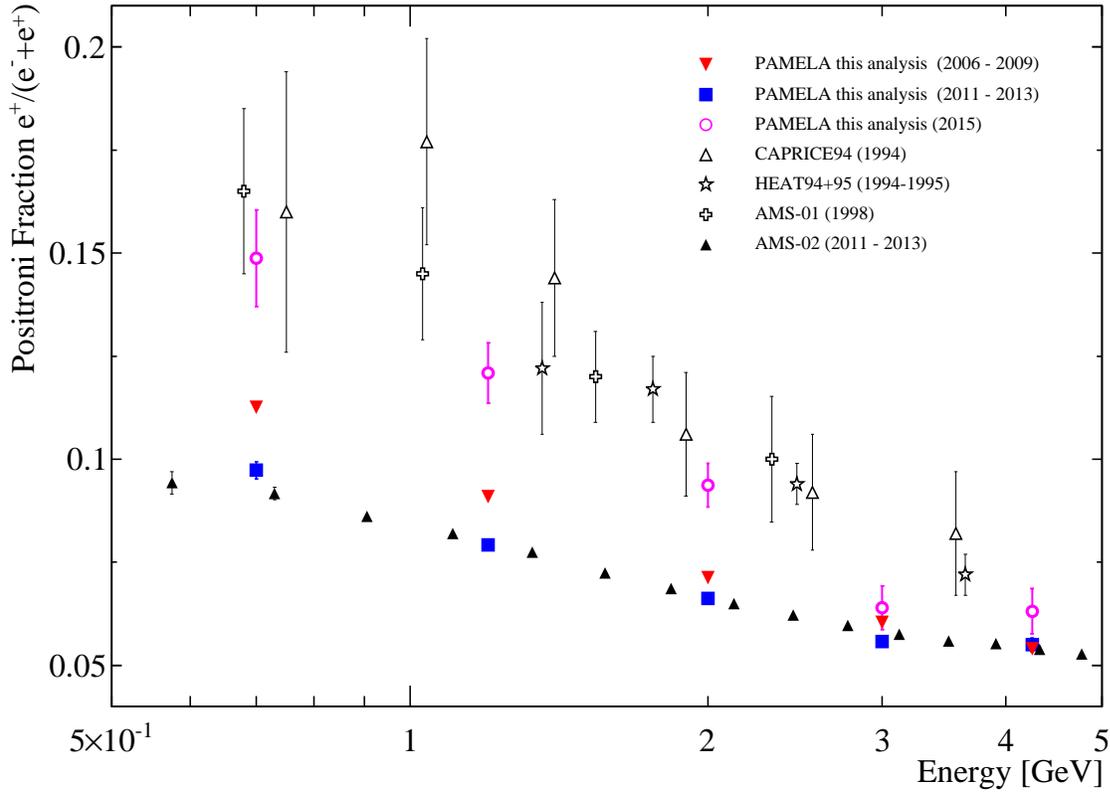}
\end{center}
\caption{The positron fraction
derived in this work for three time periods:
July 2006-December 2009, 
(solar minimum, as in \cite{adr13a}), May 2011-November 2013
(as AMS-02 results~\cite{acc14}), January-December 2015, along with
other
recent measurements: 
HEAT94+95~\protect\cite{bar97}, CAPRICE94~\protect\cite{boe00},
  AMS-01~\protect\cite{alc00b}, 
AMS-02~\protect\cite{acc14}.
The results from \cite{bar97,boe00,alc00b} refer to the previous
A$>0$ solar cycle. 
\label{f2} } 
\end{figure}
After $2009$ the solar activity started to increase and the 
CR intensity decreased up to the middle of $2013$ where it remained
constant till late $2015$. 
At the same time the ratio \ps/\el decreased until the
middle of $2012$.  
This means a stronger decrease in the positron intensity at Earth with
respect to electrons.  
Until the middle  of $2013$  the ratio remained constant and slowly
increased up to the middle of $2014$ when  
a sudden rise was observed up to late $2015$ for the first two
panels of Figure \ref{f1} where  
positrons increased respectively about $80\%$ and $50\%$ more than
electrons. This sudden rise is  
not observed for the highest energy interval, where the positrons
increased only about $20\%$ more than electrons.  
The sudden rise measured during this period appears to be a
consequence of the polarity reversal of the HMF. 

The trends in the observational data shown in Figure \ref{f1} 
can be interpreted in terms of particle drifts. 
In the context of this charge-sign dependent modulation, the tilt
angle  \cite{tilt} of the wavy heliospheric current sheet is the most
appropriate proxy  
for solar activity. For the period $2006$ to $2009$, this tilt angle
decreased slowly to reach a minimum value at the end of
$2009$. During this  
$A < 0$ magnetic polarity cycle, positrons drifted towards the Earth
mainly through the equatorial regions of the heliosphere, encountering
the  
changing wavy current sheet, while electrons drifted inwards mainly
through the polar regions of the heliosphere and were consequently less
influenced  
by the current sheet. The positron flux therefore increased relatively
more than the electron flux with a decreasing tilt angle until the end
of 
$2009$, so that the ratio $e^+/e^-$ gradually increased to the point
when solar minimum modulation conditions were settled throughout the
heliosphere.  
From $2010$ onwards, the tilt angle increased sharply so that
the positron flux also decreased proportionally faster than the
electron flux  
and the ratio $e^+/e^-$ decreased. This continued until increased
solar activity influenced both fluxes equally and the ratio $e^+/e^-$
became steady.  
From the end of $2012$, the solar magnetic field had gone into a
reversal phase, which lasted until the beginning of $2014$, when the
reversal of  
both the northern and southern solar magnetic field components was
established and the sign of the magnetic polarity in each hemisphere
became 
again clearly recognizable. After this turbulent reversal phase (from
$A < 0$ to $A > 0$) the positrons gradually started to drift inwards
through  
the polar regions of the heliosphere to the Earth while the electrons
started to drift inwards through the equatorial regions so that
the 
positron flux increased proportionally more than for
electrons. 

This can be observed also in Figure~\ref{f2} that shows
the positron fraction derived in this work for three time periods:
July 2006-December 2009 
(solar minimum, as in \cite{adr13a}), May 2011-November 2013
(as AMS-02 results~\cite{acc14}), January-December 2015, along with
previous experimental results. A good agreement between these data and
the AMS-02 results can be noticed. Moreover, the positron fraction
measured in 2015 draws near to the measurements
\cite{bar97,boe00,alc00b} from the previous
A$>0$ solar cycle in the 90's.


\paragraph{Conclusions.}
We have presented new results on the  positron and electron intensity 
below 10 GeV obtained
by the PAMELA experiment and covering the period from the  
minimum of solar cycle $23$ until the middle of the maximum
of solar 
cycle $24$, through the polarity reversal of the HMF.
Clear evidence of sign-charge dependent solar modulation
was observed. The positron fraction evolves with time as the solar
activity varies, approaching in 2015 values consistent with the
measurements from the previous A$>0$ solar cycle 22. 

\begin{acknowledgments}
This study has been partially financially supported by The Italian Space
Agency (ASI). 
We also acknowledge support from Deutsches
f\"{u}r 
Luft- und Raumfahrt (DLR), The Swedish National Space Board, The
Swedish Research 
Council, The Russian Space Agency (Roscosmos) and Russian Science
Foundation. M. Potgieter and E. Vos acknowledge the partial financial
support from the South African Research Foundation (NRF) under the
SA-Italy Bilateral Programme.  
\end{acknowledgments}

\bibliography{pamelaPsel}

\end{document}